\documentclass{article}
\usepackage{graphicx} 
\usepackage{amsmath}
\usepackage{braket}
\usepackage{bm}
\usepackage{threeparttable} 
\usepackage{url}
\usepackage{float}
\title{Free complement method with Gaussian-expanded complements: hierarchical  decontraction to mitigate  the exponential wall before selection}
\author{Cong Wang \thanks{PO Box 26 Okemos, MI, 48805 (USA); congwang.webmail@gmail.com}}
\date{}

\newcommand{\G}{\mathrm{G}}
\newcommand{\en}{\mathrm{en}}
\newcommand{\ee}{\mathrm{ee}}

\begin{document}

\maketitle

\begin{abstract}
The previous work (arXiv:2508.04635) of the free complement (FC) method with Gaussian-expanded complement functions adopts the Slater initial wavefunction. This may introduce an exponential complexity of the variational coefficients associated to the Gaussian complement functions with respect to the number of electrons at a given order before the overlap matrix based selection, for more than one Gaussian function used in the expansion. The present work uses decontractions via the distinct exponents introduced by the $g$ functions to avoid this scenario at low order of the FC method. The exponential number of the variational parameters is postponed to higher orders of the FC expansion.

\end{abstract}

\section{Introduction} \label{sec_intro}

Numerically exact solutions of quantum many-body systems are helpful in providing reference values of computational methods \cite{shiozaki2009higher,Valeev_PRF_2009_46811G6,bubin2013born,mitroy2013theory} and for testing the foundational theories \cite{bubin2013born,mitroy2013theory,PhysRevA.95.062510,henson2022measurement}. 

The FC method is a framework to compute numerically exact results for general atomic and molecular systems \cite{nakatsuji2004scaled,nakatsuji2005general,nakatsuji2012discovery}. In the previous work \cite{wang2025variational}, we have introduced the Gaussian-expanded complement function approach that all related integrals can be computed in closed forms \cite{mitroy2013theory}. This approach \cite{wang2025variational} generally avoids optimizing the exponents and potentially reduces the computational cost of the antisymmetrizations \cite{nakatsuji2015solving,nakashima2020free}.

However, for a given order of the FC method with the Gaussian complements via the expansion of the Slater-type orbital from $n$-Gaussians, (STO-$n$G) \cite{o1966gaussian,hehre1969self,hehre1970self, pietro1980molecular,pietro1981molecular,pietro1983molecular,lopez1987large,fernandez1988accurate,tew2005new, werner2007general}, $n_\G > 1$ (as in the previous work \cite{wang2025variational}, we shall use $n$G for the number of Gaussian functions in this expansion, to distinguish from order $n$ of the FC method), for the initial Slater function, this can introduce an exponential complexity in terms of the number of the complement functions with respect to the number of electrons, before the overlap selection \cite{von2008trapped,von2009correlated,rakshit2012hyperspherical,mitroy2013theory,Kalaee2014,Mosegaard2018,moriya2023novel,coomar2022quantum,wang2025variational}.

For example, consider a many-electron wavefunction for an atom as the initial wavefunction of the FC method is
\begin{align}
 \Pi_{i=1}^N e^{-\zeta_i r_i}  \label{eg_wf}
\end{align}
. Here we have omitted the spin functions, antisymmetrizations, higher angular momenta, and higher order radius functions. $\zeta_i$, $r_i$, and $N$ are the exponent coefficient of the Slater orbital, the electron position, and the number of electrons of the system, respectively. The position of the nucleus is assumed at the origin.

Using the Gaussian expansion \cite{o1966gaussian}
\begin{align}
 e^{-\zeta r}  \approx \sum_{k=1}^{n_\G} c_k e^{-\alpha_{k} r^2}   \label{gauss_expansion}
\end{align}
to expression \eqref{eg_wf} will generate $(n_\G)^N$  complement functions. Here $r$ is denoted to a general distance, including electron-nucleus  and electron-electron distances.  Since the expansion coefficients $\{  c_k  \}$ are to be determined variationally after the selection, for $n_\G >1$, this can be computationally challenging due to the exponential number of complement functions by the increasing number of electrons, even at the overlap matrix selection stage.  

This exponential complexity may be alleviated by quantum computing \cite{feynman1982simulating, aspuru2005simulated,mcardle2020quantum}, screening from bounds \cite{barca2017three,thompson2019integral}, selection scheme \cite{huron1973iterative,holmes2016heat,tubman2016deterministic,pineda2021chembot,schmerwitz2025neural,casier2026machine},  machine learning \cite{nakashima2020free,pfau2020ab,corzo2021learning,pineda2021chembot,schmerwitz2025neural,casier2026machine}, and tensor network \cite{verstraete2023density,white2023early,baiardi2020transcorrelated}.

Since the non-relativistic many-electron Hamiltonian \cite{nakatsuji2005general} (or potential by the p-alone method \cite{nakatsuji2022accurate}) and the $g$ functions \cite{nakatsuji2005general,nakatsuji2022accurate,nakatsuji2024accurate} contain one and two-electron operators and functions, for an initial wavefunction with a single complement function, it is expected that order 1 includes correlations originated from the generated one- and two-electron functions. Notice for hydrogen atom, higher-order complement functions can have further correction, if the initial wavefunction is not the exact solution \cite{nakatsuji2009does}. Therefore, order 1 can also bring corrections at one-electron level. At order 2, this would include two- to four-electron functions.

The aforementioned view suggests a possibility to construct a hierarchy of decontractions that associates to the complement functions generated by the $g$ functions. To be more specific, the decontraction is performed with the Gaussian expansion, Eq. \eqref{gauss_expansion}, of the Slater function, which has the distinct exponent introduced by the $g$ function (different from the initial wavefunction). The flexibilities of the decontracted Gaussian-expanded complement functions and the exponential-type convergence, \cite{klopper1986gaussian, kutzelnigg1994theory,kutzelnigg1996convergence,Kutzelnigg2011OWR,kutzelnigg2012rate,bakken2004expansion,mckemmish2012gaussian,kutzelnigg2013expansion,bachmayr2014error,shaw2020completeness} and Refs. \cite{kutzelnigg1996convergence,wang2013rates} for two-electron convergence,  are thus adopted in describing the associated electron correlations.  The exponential wall of the variational coefficients will be postponed to high orders.  

To this end, the present work will be organized as following. In Section \ref{methodology}, we describe the present hierarchical decontraction for the FC method with the Gaussian-expanded complement function using the helium atom ground state as the example. We  discuss strategies to lower the computational cost for many-electron systems. An alternative approaches using a single Gaussian initial wavefunction is considered. The associated algorithms are provided as modifications of the previous work \cite{wang2025variational}. In Section \ref{implementation}, the programming implementations are described. In Section \ref{results}, the numerical results of the helium ground state are provided.  In Section \ref{summary}, the summary and outlook are presented.

\section{Methodology} \label{methodology}
\subsection{Hierarchical decontractions of Gaussian-expanded complements from a Slater initial wavefunction} \label{slater_initial}
Without losing of generality, we consider the initial wavefunction \cite{nakatsuji2005general} and $g$ functions \cite{nakatsuji2022accurate,nakatsuji2024accurate} for the helium atom ground state
\begin{align}
   \psi_0 &= e^{-\zeta (r_1 + r_2 )} \label{initial_wf} \\
  g&= \{   1 - e^{-\gamma_1 r_1},    1 - e^{-\gamma_2 r_2}, \,\, 1 - e^{-\gamma_{12} r_{12}} \label{g_function_adopted} \}
\end{align}
used in the previous work \cite{wang2025variational}. $\zeta$, $\gamma_1$, $\gamma_2$, and $\gamma_{12}$ are the non-zero parameters. We omit the spin functions and set the position of the nucleus at the origin in the present work  by default.

Using the p-alone method \cite{nakatsuji2022accurate}, the general form of the FC wavefunction can be written as  \cite{nakatsuji2022accurate,nakatsuji2024accurate,wang2025variational}
\begin{align}
    ( 1 + P_{12} ) \left( g_1^{n_1}   g_2^{n_2}  g_{12}^{n_{12}}   \psi_0 \right) \label{cf_form}
\end{align}
where $n_1, n_2$, and $n_{12}$ are non-negative integers. $P_{12}$ is a permutation operator of the electrons 1 and 2. For a given order $n$ with the p-alone FC method \cite{nakatsuji2022accurate}, $n_1 + n_2 + n_{12} = n$ \cite{wang2025variational}.

In the previous work \cite{wang2025variational}, we applied the Gaussian expansion, Eq. \eqref{gauss_expansion}, to expression \eqref{cf_form} and decontracted the linear combinations of the formed Gaussian complement functions.  

In the present work, we keep the expansion coefficients $\{c_k\}$ of Eq. \eqref{gauss_expansion}, if the associated $g$ function is absent in $g_1^{n_1}   g_2^{n_2}  g_{12}^{n_{12}}   \psi_0 $ in expression \eqref{cf_form}.  We decontract the expansion, Eq. \eqref{gauss_expansion}, with the distinct exponent introduced by the $g$ function in $g_1^{n_1}   g_2^{n_2}  g_{12}^{n_{12}}   \psi_0 $ of expression \eqref{cf_form}. For example, $n_1 = 1, n_2 = 0, n_{12} = 0$ in expression \eqref{cf_form} will be converted into
\begin{align}
 & ( 1 + P_{12} ) \left[ e^{-\alpha_1^{(1)} r_1^2} \left( \sum_{k=1}^{n_\G} c_2^{(k)} e^{-\alpha_2^{(k)} r_2^2} \right) \right] \label{010_1}  \\
 & ( 1 + P_{12} ) \left[   e^{-\alpha_1^{(2)} r_1^2}  \left(\sum_{k=1}^{n_\G} c_2^{(k)} e^{-\alpha_2^{(k)} r_2^2}  \right) \right] \label{010_2}  \\
  & \cdots \nonumber \\
 &( 1 + P_{12} )  \left[  e^{-\alpha_1^{(n_\G)} r_1^2}  \left(\sum_{k=1}^{n_\G} 
 c_2^{(k)} e^{-\alpha_2^{(k)} r_2^2}  \right) \right] \label{010_12}
\end{align}
Here $e^{-\alpha_1^{(1)}}, e^{-\alpha_1^{(2)}}, \cdots, e^{-\alpha_1^{(n_\G)}} $ come from the decontracted Gaussian expansion, Eq. \eqref{gauss_expansion}, of $e^{-(\zeta + \gamma_1) r_1}$.  Since $\gamma_1 \neq 0$, $\zeta+\gamma_1$ differs from the exponent, $\zeta$, of the initial wavefunction, Eq. \eqref{initial_wf}. 
$\{ c_2^{(k)} \}$ and $\{ \alpha_2^{(k)} \}$ come from the Gaussian expansion, Eq. \eqref{gauss_expansion}, of $e^{-\zeta r_2}$ without decontraction, since the exponent $\zeta$ is the same as the initial wavefunction, Eq.  \eqref{initial_wf}.

The choice between contraction and decontraction is made before the symmetrization of spatial wavefunction (manifested by the antisymmetrization of the total wavefunction), since the antisymmetrizations may be  treated approximately at large distances especially for more electron systems \cite{nakatsuji2015solving}.

In addition, the exponents associated to each variable, $r_1, r_2$, or $r_{12}$, of the Slater functions are monotonously increasing with respect to enlarging the order of the present FC method, e.g., $n \gamma_1, n = 0, 1, 2, \cdots$. For a fixed number of $n_\G$, the scaling relation \cite{o1966gaussian} 
\begin{align}
   \frac{\zeta'}{\zeta} = \sqrt{\frac{\alpha'}{\alpha}} \label{scaling}
\end{align}
will introduce a non-zero lower bound of the expanded Gaussian exponents $\{ \alpha_k \}$. This may cause describing the long-range behavior of the exact ground wavefunction and density \cite{morrell1975calculation,hoffmann1977schrodinger,hoffmann1979lower,hoffmann1980lower,katriel1980asymptotic,ahlrichs1981bounds,simon1982schrodinger,froese1983exponential,ahlrichs1989basic,fournais2008local,agmon2014lectures} less efficiently.

On the other hand, increasing $n_\G$ in the STO-$n$G basis set turns to cover smaller Gaussian exponents $\{\alpha_k \}$ \cite{o1966gaussian,hehre1969self,hehre1970self, pietro1980molecular,pietro1981molecular,pietro1983molecular,lopez1987large,fernandez1988accurate,tew2005new, werner2007general} and leads to effective convergence of the ground state energy \cite{wang2025variational}. This aligns with one sufficient condition of the energy convergence, including the exponents approaching to zero \cite[Theorem 5]{klahn1977convergence}.

Notice the present approach is for reducing the number of the variational coefficients. If one uses the integration formulae of explicitly correlated Gaussian functions \cite{mitroy2013theory} including the non-geminal functions, the number of integrals are expected to be comparable with the previous approach \cite{wang2025variational} before the overlap selection \cite{von2008trapped,von2009correlated,rakshit2012hyperspherical,mitroy2013theory,Kalaee2014,Mosegaard2018,moriya2023novel,coomar2022quantum,wang2025variational}. However, we can think the Slater function at the left hand side of the Gaussian expansion, Eq. \eqref{gauss_expansion}, as the input of the L\"{o}wdin rules \cite{lowdin1955quantum_a,lowdin1955quantum_b} and the generalization \cite{nakashima2020free}. Up to the complement functions at the levels  \cite{nakashima2020free} of one geminal function, the associated integrals can be expressed within the four-electron integrals and the minors with  polynomial computational time \cite{lowdin1955quantum_a,lowdin1955quantum_b,nakashima2020free}.

For example, the overlap integral between two Slater determinants can be expressed as an $N \times N$ determinant of the overlap matrices of the one-electron functions \cite{lowdin1955quantum_a,lowdin1955quantum_b}. One element of the resulting determinant
can be computed as 
\begin{align}
  & \int d^3 \bm{r}_i (e^{-\zeta_i r_i})^*e^{-\zeta_i r_i} \label{int_product_0} \\
  &= \sum_{k_i k_i'=1}^{n_\G}  \int d^3 \bm{r}_i c_{k_i}^* c_{k'_i}  (e^{-\alpha_i^{(k)} r_i^2})^* e^{-\alpha^{(k')}_i r^2_i} \label{int_product}
\end{align}
Here $^*$ denotes to complex conjugation. The computational costs of Eq. \eqref{int_product}, all elements in the resulting determinant, and the subsequent evaluation of the resulting determinant will be $O((n_\G)^2)$, $O(N^2 (n_\G)^2)$, and $O(N^3)$, respectively.

For sparse systems, the antisymmetrizations can be approximately treated within a certain accuracy \cite{nakatsuji2015solving}. The overlap integrals would resemble the Hartree product and its approximate permutations. For example,
\begin{align}
  & \Pi_{i=1}^N \int \cdots \int d^3 \bm{r}_i (e^{-\zeta_i r_i})^*e^{-\zeta_i r_i} \label{int_product_hartree_0} \\
  &= \Pi_{i=1}^N \sum_{k_i k_i'=1}^{n_\G}  \int \cdots \int d^3 \bm{r}_i c_{k_i}^* c_{k'_i}  (e^{-\alpha_i^{(k)} r_i^2})^* e^{-\alpha^{(k')}_i r^2_i} \label{int_product_hartree}
\end{align}
One can evaluate the integrals and summation of each electron, then multiply the result of each electron. The cost is $O(N (n_\G)^2)$. The many-electron Hamiltonian and geminal functions from low-order FC expansions would be expected to introduce polynomial modifications of Eq. \eqref{int_product_hartree}.

The four-electron integrals from the generalization of the L\"{o}wdin rules  \cite{nakashima2020free}  would involve 8-fold loops over a number of $n_\G$ Gaussian expansion of the one-electron functions, when the $g$ functions only appear as the geminal functions (not the one-electron functions) for order 1 or the $s_{ij}$-assisted approach  \cite{nakashima2020free} in higher orders of the geminal functions. This can be computationally demanding. 

As will be shown in the Subsection \ref{energy_sel}, the preliminary result of energy-based selection \cite{nakashima2020free,nakatsuji2020solving} of the helium ground state indicates fewer complement functions would be needed with the presence of geminal functions, than expanding the initial wavefunction within a certain accuracy. This suggests we may use fewer number of the Gaussian expansions with the geminal complements, e.g., choosing a small number of $n_\G$ or discard some exponents and contraction coefficients (set as zero) of the $n_\G$ expansion. A reduction of computation cost  may be anticipated.

Additionally, tensor decomposition techniques can reduce the folds of the loops. For example \cite{Limpanuparb2011,parrish2013exact,khoromskaia2015tensor,khoromskaia2018tensor,Glaser2024}
\begin{align}
   \frac{1}{r_{13}} \approx \sum_a f_a (\bm{r}_1) g_a (\bm{r}_3) 
\end{align}
would approximate  $\langle e^{-\alpha_{12} r_{12}^2} r_{13}^{-1} e^{-\alpha_{34} r_{34}^2} \rangle $ into 5-fold summations.

The tensor decomposition will introduce approximations of the matrix elements. Thus, the computation would no longer be variational \cite{kutzelnigg1991wave}. Nevertheless, if the introduced errors are sufficiently small \cite{kutzelnigg1991wave}, these may not significantly affect the second-order error \cite{helgaker2000molecular} of the variational method.

\subsection{Single Gaussian initial wavefunction} \label{gauss_initial}

If the initial wavefunction is a Gaussian function
\begin{align}
  \psi_0 = e^{-\alpha (r_1^2 + r_2^2 )} \label{initial_wf_gto}   
\end{align}
this could automatically furnish the hierarchical number of Gaussian-expanded complements, by decontracting the Gaussian expansion, Eq. \eqref{gauss_expansion}, and within the p-alone method \cite{nakatsuji2022accurate} as in the previous work \cite{wang2025variational}. The Slater functions from the $g$ functions, Eq. \eqref{g_function_adopted}, combine with the initial wavefunction, Eq. \eqref{initial_wf_gto}. After the Gaussian expansion, Eq. \eqref{gauss_expansion}, it includes
\begin{align}
 e^{-\alpha_1^{(k)} r_1^2 } e^{-\alpha (r_1^2 + r_2^2 )} , k \in 1, \cdots, n_\G
\end{align}
Thus, $\alpha$ in the initial Gaussian wavefunction, Eq. \eqref{initial_wf_gto},  becomes a lower bound in all Gaussian exponents of $r_1$ or $r_2$.

To the contrary of the Slater initial wavefunction in Subsection \ref{slater_initial}, this lower bound cannot be reduced by enlarging $n_\G$ in expanding the initial Slater function,  Eq. \eqref{initial_wf}.
The results in Subsection \ref{gto_result} suggest, for the tested $\gamma$ parameters in the $g$ functions, this approach is computationally inefficient.

\subsection{Algorithms} \label{alg}
The routes correspond to Subsections \ref{slater_initial} and \ref{gauss_initial} are similar to Algorithms 1 and 2 in the  previous work \cite{wang2025variational}.  We describe the present approach as modifications of Algorithms 1 and 2 of Ref. \cite{wang2025variational}.

For the Slater initial wavefunction with hierarchical Gaussian expansion in Subsection \ref{slater_initial}, the formation of the Cartesian product in step 7 of the Algorithm 1, in the term absent with the $g$ function, we use the exponents and coefficients as one element to form the Cartesian product in the present approach. The example in Eqs. \eqref{010_1} - \eqref{010_12} corresponds to 
\begin{align}
\left[   [[\alpha_1^{(1)},  \alpha_1^{(2)}, \alpha_1^{(3)} ], \right. &  \left. [c_1^{(1)},  c_1^{(2)}, c_{1}^{(3)} ]]   \right]  \label{fold_1}  \\
 &\times   \nonumber \\
  [\alpha_2^{(1)},  &\alpha_2^{(2)}, \alpha_2^{(3)} ]  \label{fold_2}  \\
   &\times \nonumber \\
   [[0], & [1]]  \label{fold_12}  \\
  =\left(  [[\alpha_1^{(1)},  \alpha_1^{(2)}, \alpha_1^{(3)} ],  \right. & [c_1^{(1)},  c_1^{(2)}, c_{1}^{(3)} ] ],  \left. \alpha_2^{(1)},  [[0],[1]] \right), \label{cartesian_prod_1} \\
  \left(  [[\alpha_1^{(1)},  \alpha_1^{(2)}, \alpha_1^{(3)} ], \right. & \left.  [c_1^{(1)},  c_1^{(2)}, c_{1}^{(3)} ] ], \alpha_2^{(2)}, [[0],[1]] \right),  \label{cartesian_prod_2}  \\
   \left(  [[\alpha_1^{(1)},  \alpha_1^{(2)}, \alpha_1^{(3)} ],  \right. & \left. [c_1^{(1)},  c_1^{(2)}, c_{1}^{(3)} ] ], \alpha_2^{(3)},  [[0],[1]] \right)  \label{cartesian_prod_12}
\end{align}
Expressions \eqref{fold_1}, \eqref{fold_2}, and \eqref{fold_12} are contracted, decontracted, and contracted, respectively. The contraction is with the exponent in the initial wavefunction Eq. \eqref{initial_wf}, $\zeta$ or $0$ with $r_2$ or $r_{12}$, respectively. The decontraction is with the distinct exponent introduced by the $g$ function, here, $\zeta + \gamma_1$ with $r_1$ as in Subsection \ref{slater_initial}. Contracted entry returns exponents and coefficients. Decontracted entries return exponents.

Notice $e^{0 r_{12}}$ of the STO-$n$G expansion in the present work leads one exponent 0 and one coefficient 1, as in expression \eqref{fold_12}. In the previous work \cite{wang2025variational}, all entries are decontracted and  $e^{0 r_{12}}$ returns a single zero exponent. 

In addition, the linear expansion coefficients of the STO-$n$G basis sets in Refs. \cite{o1966gaussian,hehre1969self} were reported for the normalized Slater functions with the exponent $\zeta = 1$ and normalized Gaussian functions. Namely \cite{o1966gaussian,hehre1969self},
\begin{align}
    \sqrt{\frac{1}{\pi}} e^{-r} \approx \sum_{k=1}^{n_\G} c_k  \left(\frac{2\alpha_k}{\pi} \right)^{3/4} e^{-\alpha_k r^2} \label{norm_linear_coeffs}
\end{align}
In Ref. \cite{fernandez1988accurate}, the linear expansion coefficients were reported as in Eq. \eqref{gauss_expansion}. In the present formulations with Eq. \eqref{gauss_expansion}, the coefficient $c_k$ via Eq. \eqref{norm_linear_coeffs} of Refs. \cite{o1966gaussian,hehre1969self}  is multiplied by $\sqrt{\pi} ( 2\alpha_k / \pi)^{3/4} $. 

In step 8 of Algorithm 1, the sorting of the exponents of $r_1$ and $r_2$ is made only for the decontracted exponents. The contracted exponents are sorted by construction in step 3 of Algorithm 1. This also avoids complexities in determining numerically the same for both coefficients and exponents in the contracted entries.

In step 8 of Algorithm 1, both the previous \cite{wang2025variational} and present work have regarded two exponents in a decontracted complement function are the same and without sorting, if the relative error with respect to the larger one in absolute value is within $1\times 10^{-9}$ \cite{python_math}. Whether adopting this setting is not expected to lead to a numerical difference, since the thresholds of the overlap selection are 0.95 - 0.9995 (similar to the values in Ref. \cite{mitroy2013theory}).

After step 10 of Algorithm 1, a class type is formed. Along the example with Eqs. \eqref{cartesian_prod_1} - \eqref{cartesian_prod_12}, they are converted into three entries. Each one has \texttt{exps\_coeffs} and \texttt{linear\_coeffs} elements:
\begin{align}
\texttt{exps\_coeffs}: [ [\alpha_1^{(1)},  \alpha_2^{(1)}, 0 ], [\alpha_1^{(2)},  \alpha_2^{(1)}, 0 ], [\alpha_1^{(3)},  \alpha_2^{(1)}, 0 ]], \texttt{linear\_coeffs}:  [c_1^{(1)}, c_1^{(2)}, c_1^{(3)}]  \label{return_1} \\
\texttt{exps\_coeffs}: [ [\alpha_1^{(1)},  \alpha_2^{(2)}, 0 ], [\alpha_1^{(2)},  \alpha_2^{(2)}, 0 ], [\alpha_1^{(3)},  \alpha_2^{(2)}, 0 ]], \texttt{linear\_coeffs}:  [c_1^{(1)}, c_1^{(2)}, c_1^{(3)}]  \label{return_2}   \\
\texttt{exps\_coeffs}: [ [\alpha_1^{(1)},  \alpha_2^{(3)}, 0 ], [\alpha_1^{(2)},  \alpha_2^{(3)}, 0 ], [\alpha_1^{(3)},  \alpha_2^{(3)}, 0 ]], \texttt{linear\_coeffs}:  [c_1^{(1)}, c_1^{(2)}, c_1^{(3)}]   \label{return_3}   
\end{align}
They are used in the return entries of Algorithm 1. \texttt{linear\_coeffs} is formed by the multiplications of the coefficients in the Cartesian product. For example, in expression \eqref{cartesian_prod_1}, $c_1^{(1)} \cdot 1 \cdot 1 = c_1^{(1)}$ which pairs with $[\alpha_1^{(1)}, \alpha_2^{(1)}, 0]$. Here the coefficient 1 is assigned to the decontracted entry $\alpha_2^{(1)}$ in expression \eqref{cartesian_prod_1}.

In steps 9 and 10 of Algorithm 2, we have removed the $1 \times 10^{-12}$ threshold of the exponent difference with $r_1$ and $r_2$ for symmetrizing spatial functions \cite{wang2025variational}. All spatial functions are symmetrized. This can avoid comparing both coefficients and exponents in contracted entries.

In step 11 of Algorithm 2 of both the previous \cite{wang2025variational} and the present work, we loop over complement functions before the overlap selection, in terms of $i=1 \,\, \textrm{to} \,\, M_n^{\textrm{before}}$. If the complement function is not in $\texttt{duplicated gfcs}$, it is added in the list for return of Algorithm 2 after the overlap selection. Hence, it forms a sequence of complement functions after the overlap selection. Only the complement functions after the overlap selection was symmetrized in symbolic form in step 12 of Algorithm 2 of the previous work \cite{wang2025variational}.

In steps 12 and 13 of Algorithm 2, no symmetrization nor forming symbolic form of complement functions is made in the present work. The returned complement functions of Algorithm 2 are in the form of expressions \eqref{return_1} - \eqref{return_3}. The symmetrizations in the complement functions are made in forming the Hamiltonian and overlap matrices $\mathbf{H}$ and $\mathbf{S}$ respectively, in the subsequent secular equation, $\mathbf H \mathbf C = \mathbf S \mathbf C \mathbf E.$

In the present work, we use the formulae in Ref. \cite{mitroy2013theory} to compute the integrals without the L\"owdin rules \cite{lowdin1955quantum_a,lowdin1955quantum_b}. We incorporate the 16-fold symmetries of the overlap, sum of electron-nuclei attractions, and electron-electron repulsion integrals and 4-fold symmetries of the kinetic integrals. Namely, denoting 
\begin{align}
  I_{a_1a_2a_{12}b_1b_2b_{12} } (O) := \braket{e^{-a_1 r_1^2 -a_2 r_2^2 -a_{12} r_{12}^2}| \hat O |e^{-b_1 r_1^2 -b_2 r_2^2 -b_{12} r_{12}^2}}  \label{int_symm}
\end{align}
, here no symmetrization in the ket and bra parts on the right-hand side of Eq. \eqref{int_symm}, we have 
\begin{align}
  I_{a_1a_2a_{12}b_1b_2b_{12} } (O) &=   I_{b_1a_2a_{12}a_1 b_2b_{12} } (O), a_1 \leftrightarrow b_1 \\
 &=   I_{a_1b_2a_{12}b_1a_2b_{12} } (O) , a_2 \leftrightarrow b_2 \\
  &=   I_{a_1a_2b_{12}a_1 a_2a_{12} } (O) , a_{12} \leftrightarrow b_{12}  \\
   &=   I_{a_2a_1a_{12}b_2b_1b_{12} } (O), 1 \leftrightarrow 2 \, ( a_1 \leftrightarrow a_2,   b_1 \leftrightarrow b_2)
\end{align}
for $\hat O = \hat 1, \hat V_{\en,1} + \hat V_{\en,2}, \hat V_{\ee}  $ 
\begin{align}
  I_{a_1a_2a_{12}b_1b_2b_{12} } (O) &=   I_{b_1b_2b_{12}a_1 a_2 a_{12} } (O), a\leftrightarrow b \, (a_1 \leftrightarrow b_1, a_2 \leftrightarrow b_2, a_{12} \leftrightarrow b_{12}) \text{\cite{silvestre2007few}} \\
 &=   I_{a_2a_1a_{12}b_2b_1b_{12} } (O) , 1 \leftrightarrow 2 \, ( a_1 \leftrightarrow a_2,   b_1 \leftrightarrow b_2) 
\end{align}
for $\hat O =  \hat T_1 + \hat T_2 $. The combinations of these permutations cover the aforementioned 16- and 4-fold symmetries.

To realize these symmetries in the overlap selection, forming the Hamiltonian and overlap matrices, and loading the recomputed integrals, hash tables \cite{garniron2019quantum} for the overlap, kinetic, electron-nucleus attraction, and electron-electron repulsion integrals are formed. In each hash table, the key, ($a_1, a_2, a_{12}, b_1, b_2, b_{12}$), is the minimal lexicographical value among all symmetry equivariant exponents. The value is the numerical result of the associated integral. 

To reduce the timing of generating all symmetry equivalent exponents, additional hash tables for the 16-fold and 4-fold symmetry exponents are introduced. In each of these additional hash tables, the key is the input exponents. The value is the lexicographically minimal  one among all symmetry equivalent exponents.

\section{Implementations} \label{implementation}
Similar to the previous work \cite{wang2025variational}, we have used the \texttt{Python} program language \cite{van1991interactively} version 3.12.3 with \texttt{mpmath} multi-precision library version 1.3.0 \cite{mpmath} for the numerical computations, except for the \texttt{time} \cite{time_python} function and plotting with  the \texttt{matplotlib} library \cite{Hunter:2007} version 3.8.0. \texttt{SymPy} library version 1.14.0 \cite{10.7717/peerj-cs.103} and \texttt{cachetools} library version 7.0.3 \cite{cachetools} were used for the symbolic computations and memory management of integral storage, respectively. The integral formulae in Ref. \cite{mitroy2013theory} are implemented for two-electron systems (also in the previous works \cite{wang2025variational}). The standard libraries, \texttt{itertools} and \texttt{math} for forming Cartesian product between exponents of $r_1, r_2, r_{12}$ and for numerical comparisons, respectively, were used.  The \texttt{Julia} program language \cite{Julia-2017} version 1.11.6 with \texttt{GenericLinearAlgebra} package version 0.3.18 and \texttt{BigFloat} for multi-precision computations used in the previous work was adopted \cite{wang2025variational}.  The Julia library JSON \cite{json} version 0.21.4 was used in data transfer between Python and Julia (also in the previous works \cite{wang2025variational}). 

Both \texttt{mpmath} and \texttt{BigFloat} are set at 50 decimal precision. We have used the canonical orthogonalization scheme as in the previous work, where the matrices associated to eigenvalues of the overlap matrix below $1\times 10^{-30}$ will be discarded \cite{wang2025variational}. Nevertheless, the eigenvalues of the overlap matrices are all above this threshold in the present work. 

For Figure \ref{fig1}, double precision is used.

In using step 1 of Algorithm 2 as in Ref. \cite{wang2025variational}, \texttt{duplicated gfcs} is in set type to enable the hash search efficiency over the list type \cite{python_set}.

For the hash table in loading integral values, all indices in the key are in double-precision format for efficiency. This lower than the \texttt{mpmath} 50 decimal precision is not expected to cause a numerical difference for the ground state energy, if assigning two indices differing below double precision into a single entry. The thresholds of overlap selection, 0.95 - 0.9995, in the present work may exclude  sufficiently numerically close exponents.

The \texttt{cursor} developing environment with large language models \cite{cursor} has been used in generating the \texttt{Python} code.

\section{Numerical results and discussions} \label{results}
\subsection{Hierarchical decontraction with Slater initial wavefunction}

In this subsection, we present numerical results for the Gaussian expansion on the complement functions as products of the $g$ functions Eq. \eqref{g_function_adopted} in the $p$-alone method \cite{nakatsuji2022accurate} with the Slater initial wavefunction Eq. \eqref{initial_wf} for the helium ground state. $\zeta = 1.6875$ in Eq. \eqref{initial_wf}, $\gamma_1 = \gamma_2 = 0.3125, \gamma_{12} = 0.5 $ in Eq. \eqref{g_function_adopted} are chosen according to Ref. \cite{nakatsuji2022accurate}.  The atomic units (a.u.) are used by default in the present work.

The computational data are provided in Tables \ref{tab1} and \ref{tab2}. Subchemical accuracy (0.1 kcal mol$^{-1}$)  of the total energy with respect to the reference values \cite{schwartz2006further,nakashima2007solving,kurokawa2008solving,nakashima2008accurately} can be obtained at $n=2, n_\G=14$ or $n=3, n_\G=10$ with 0.95 overlap matrix selection threshold or $n=2, n_\G=10$ with 0.98 overlap matrix selection threshold. These are at higher levels, in terms of $n, n_\G$, and overlap selection threshold, than the decontracted approach at  $n=2, n_\G=10$ with 0.95 overlap selection threshold \cite{wang2025variational}. This is possibly due to the constraints from the contractions. 

Total energy error with respect to the reference values \cite{schwartz2006further,nakashima2007solving,kurokawa2008solving,nakashima2008accurately} less than $1\times 10^{-6}$ a.u. can be reached at  $n=3, n_\G=14$ or $n=3, n_\G=10$ with 0.99 overlap matrix selection threshold or $n=3, n_\G=10$ with 0.995 overlap matrix selection threshold.

\begin{table}[H]
\caption{Results of the variational calculations via the Gaussian-expanded complement functions for the ground state of the helium atom. $\psi_0 = e^{-\zeta (r_1 + r_2) }$, $\zeta = 1.6875 $ \cite{nakatsuji2022accurate}. 0.95 and 0.98 are the overlap thresholds in Algorithm 2 of Ref. \cite{wang2025variational}, that are used in screening the linear dependent Gaussian complement functions. The column  $n_\G$ corresponds to the decontracted STO-$n$G basis set \cite{hesselmann2005efficient} ($n_\G = 3,6$ \cite{hehre1969self}, $10$ \cite{o1966gaussian}, and $14$ \cite{fernandez1988accurate}). $M_n^{\mathrm{before}}$ and $M_n^{\mathrm{after}}$ are denoted to the number of the complement functions before and after the selection of the values of overlap matrices, respectively.  $s_{\mathrm{min}}$ stands for the minimum eigenvalue of the overlap matrix after the selection (also for the previous work \cite{wang2025variational}). $E$ stands for the electronic energy \cite[p. 43]{szabo1996modern}. Energies and $s_{\mathrm{min}}$ are rounded to even \cite{python_format,python_round,goldberg1991every}. }

\begin{threeparttable}
\begin{tabular}{ccccccccc}
\hline
\multicolumn{3}{c}{} & \multicolumn{3}{c}{0.95} & \multicolumn{3}{c}{0.98} \\
$n$ &  $n_\G$ &  $M_n^{\mathrm{before}}$ &  $M_n^{\mathrm{after}}$  &  $s_{\mathrm{min}}$ & $ E$ & $M_n^{\mathrm{after}}$  & $s_{\mathrm{min}}$  &  $ E$   \\ \hline
0 & 3 & 1 & 1 & 1.0 $\times$ 10$^{0}$ & -2.807801 & 1 & 1.0 $\times$ 10$^{0}$ & -2.807801 \\
 & 6 & 1 & 1 & 1.0 $\times$ 10$^{0}$ & -2.846299 & 1 & 1.0 $\times$ 10$^{0}$ & -2.846299 \\
 & 10 & 1 & 1 & 1.0 $\times$ 10$^{0}$ & -2.847646 & 1 & 1.0 $\times$ 10$^{0}$ & -2.847646 \\
 & 14 & 1 & 1 & 1.0 $\times$ 10$^{0}$ & -2.847655 & 1 & 1.0 $\times$ 10$^{0}$ & -2.847655 \\
1 & 3 & 7 & 3 & 1.3 $\times$ 10$^{-1}$ & -2.851123 & 5 & 5.3 $\times$ 10$^{-4}$ & -2.858924 \\
 & 6 & 13 & 8 & 4.2 $\times$ 10$^{-3}$ & -2.895993 & 9 & 1.3 $\times$ 10$^{-3}$ & -2.897547 \\
 & 10 & 21 & 15 & 1.8 $\times$ 10$^{-3}$ & -2.899259 & 16 & 4.2 $\times$ 10$^{-4}$ & -2.899427 \\
 & 14 & 29 & 18 & 8.3 $\times$ 10$^{-4}$ & -2.897938 & 22 & 7.4 $\times$ 10$^{-6}$ & -2.899441 \\
2 & 3 & 28 & 9 & 1.6 $\times$ 10$^{-3}$ & -2.858493 & 14 & 6.4 $\times$ 10$^{-6}$ & -2.867025 \\
 & 6 & 82 & 32 & 4.2 $\times$ 10$^{-5}$ & -2.901822 & 41 & 3.8 $\times$ 10$^{-6}$ & -2.902180 \\
 & 10 & 196 & 81 & 3.4 $\times$ 10$^{-5}$ & -2.903557 & 104 & 4.2 $\times$ 10$^{-8}$ & -2.903692 \\
 & 14 & 358 & 129 & 3.6 $\times$ 10$^{-6}$ & -2.903591 & 180 & 1.9 $\times$ 10$^{-10}$ & -2.903704 \\
3 & 3 & 79 & 13 & 3.7 $\times$ 10$^{-4}$ & -2.860757 & 24 & 3.3 $\times$ 10$^{-6}$ & -2.889824 \\
 & 6 & 328 & 47 & 3.2 $\times$ 10$^{-5}$ & -2.902878 & 82 & 2.6 $\times$ 10$^{-7}$ & -2.903103 \\
 & 10 & 1066 & 130 & 3.8 $\times$ 10$^{-6}$ & -2.903664 & 218 & 2.0 $\times$ 10$^{-8}$ & -2.903720 \\
 & 14 & 2444 & 182 & 1.6 $\times$ 10$^{-6}$ & -2.903705 & 338 & 1.6 $\times$ 10$^{-10}$ & -2.903721 \\
\hline
\end{tabular} \label{tab1}
\end{threeparttable}
\end{table}

\begin{table}[H]
\caption{Results of the variational calculations via the Gaussian-expanded complement functions for the ground state of the helium atom. $\psi_0 = e^{-\zeta (r_1 + r_2) }$, $\zeta = 1.6875 $ \cite{nakatsuji2022accurate}. 0.99 and 0.995 are the overlap thresholds in Algorithm 2 of Ref. \cite{wang2025variational}, that are used in screening the linear dependent Gaussian complement functions. The column  $n_\G$ corresponds to the decontracted STO-$n$G basis set ($n_\G= 3,6$ \cite{hehre1969self}, $10$ \cite{o1966gaussian}, and $14$ \cite{fernandez1988accurate}). $M_n^{\mathrm{before}}$ and $M_n^{\mathrm{after}}$ are denoted to the number of the complement functions before and after the selection of the values of overlap matrices, respectively.  $s_{\mathrm{min}}$ stands for the minimum eigenvalue of the overlap matrix after the selection. $E$ stands for the electronic energy \cite[p. 43]{szabo1996modern}. Energies and $s_{\mathrm{min}}$ are rounded to even \cite{python_format,python_round,goldberg1991every}. }

\begin{threeparttable}
\begin{tabular}{ccccccccc}
\hline
\multicolumn{3}{c}{} & \multicolumn{3}{c}{0.99} & \multicolumn{3}{c}{0.995} \\
$n$ &  $n_\G$ &  $M_n^{\mathrm{before}}$ &  $M_n^{\mathrm{after}}$  &  $s_{\mathrm{min}}$ & $ E$ & $M_n^{\mathrm{after}}$  & $s_{\mathrm{min}}$  &  $ E$   \\ \hline
0 & 3 & 1 & 1 & 1.0 $\times$ 10$^{0}$ & -2.807801 & 1 & 1.0 $\times$ 10$^{0}$ & -2.807801 \\
 & 6 & 1 & 1 & 1.0 $\times$ 10$^{0}$ & -2.846299 & 1 & 1.0 $\times$ 10$^{0}$ & -2.846299 \\
 & 10 & 1 & 1 & 1.0 $\times$ 10$^{0}$ & -2.847646 & 1 & 1.0 $\times$ 10$^{0}$ & -2.847646 \\
 & 14 & 1 & 1 & 1.0 $\times$ 10$^{0}$ & -2.847655 & 1 & 1.0 $\times$ 10$^{0}$ & -2.847655 \\
1 & 3 & 7 & 6 & 4.2 $\times$ 10$^{-4}$ & -2.860535 & 6 & 4.2 $\times$ 10$^{-4}$ & -2.860535 \\
 & 6 & 13 & 11 & 1.2 $\times$ 10$^{-5}$ & -2.897601 & 11 & 1.2 $\times$ 10$^{-5}$ & -2.897601 \\
 & 10 & 21 & 18 & 1.2 $\times$ 10$^{-7}$ & -2.899433 & 18 & 1.2 $\times$ 10$^{-7}$ & -2.899433 \\
 & 14 & 29 & 24 & 9.2 $\times$ 10$^{-9}$ & -2.899441 & 25 & 9.2 $\times$ 10$^{-9}$ & -2.899462 \\
2 & 3 & 28 & 18 & 7.8 $\times$ 10$^{-7}$ & -2.875422 & 23 & 7.7 $\times$ 10$^{-7}$ & -2.888365 \\
 & 6 & 82 & 56 & 4.8 $\times$ 10$^{-10}$ & -2.903112 & 64 & 4.8 $\times$ 10$^{-10}$ & -2.903131 \\
 & 10 & 196 & 133 & 5.6 $\times$ 10$^{-13}$ & -2.903705 & 155 & 5.5 $\times$ 10$^{-14}$ & -2.903710 \\
 & 14 & 358 & 233 & 1.9 $\times$ 10$^{-14}$ & -2.903713 & 270 & 2.0 $\times$ 10$^{-15}$ & -2.903714 \\
3 & 3 & 79 & 38 & 7.6 $\times$ 10$^{-8}$ & -2.893412 & 53 & 1.5 $\times$ 10$^{-8}$ & -2.894797 \\
 & 6 & 328 & 139 & 3.9 $\times$ 10$^{-10}$ & -2.903429 & 183 & 1.5 $\times$ 10$^{-11}$ & -2.903444 \\
 & 10 & 1066 & 386 & 6.4 $\times$ 10$^{-14}$ & -2.903721 & 494 & 3.6 $\times$ 10$^{-14}$ & -2.903723\tnote{a} \\
 & 14 & 2444 & 663 & 1.6 $\times$ 10$^{-15}$ & -2.903724\tnote{b} & 891 & 2.6 $\times$ 10$^{-16}$ & -2.903724\tnote{c} \\
\hline
\end{tabular} \label{tab2}
    \begin{tablenotes}
      \item[a] -2.903723449.
      \item[b] -2.903723973.
      \item[c] -2.903724097.    
    \end{tablenotes}
\end{threeparttable}
\end{table}

\subsection{Decontraction with a single Gaussian initial wavefunction} \label{gto_result}

In Table \ref{tab3}, the data of the variational calculations using a single Gaussian initial wavefunction, Eq. \eqref{initial_wf_gto}, with the $g$ functions, Eq. \eqref{g_function_adopted}, in the p-alone method \cite{nakatsuji2022accurate} are presented. $\gamma_1 = \gamma_2 = 2.0$ since Gaussian functions in Eq. \eqref{initial_wf_gto} have no contribution to the cusp value \cite{kato1957eigenfunctions,pack1966cusp}. $\gamma_{12} = 0.5$ are used as in Ref. \cite{nakatsuji2022accurate}. 

Table \ref{tab3} starts from the order $n=1$, since at the order $n=0$ no Gaussian expansion, Eq. \eqref{gauss_expansion}, in the initial wavefunction, Eq. \eqref{initial_wf_gto}. The energy of the order 0 is known as -2.301 \cite{nakatsuji2005general}.

The results in Table \ref{tab3} suggest, with the present parameters of the $g$ functions, this approach is computationally inefficient.

\begin{table}[H]
\caption{Results of the variational calculations via Gaussian-expanded complement functions for the ground state of the helium atom. $\psi_0 = e^{-\alpha (r_1^2 + r_2^2) }$, $\alpha = 0.767 $ \cite{nakatsuji2022accurate}. 0.995 and 0.9995 are the overlap thresholds in Algorithm 2 of Ref. \cite{wang2025variational}, that are used in screening the linear dependent Gaussian complement functions. The column  $n_\G$ corresponds to the decontracted STO-$n$G basis set ($n=10$ \cite{o1966gaussian} and $n=14$ \cite{fernandez1988accurate}). $M_n^{\mathrm{before}}$ and $M_n^{\mathrm{after}}$ are denoted to the number of the complement functions before and after the selection of the values of overlap matrices, respectively.  $s_{\mathrm{min}}$ stands for the minimum eigenvalue of the overlap matrix after the selection. $E$ stands for the electronic energy \cite[p. 43]{szabo1996modern}. Energies and $s_{\mathrm{min}}$ are rounded to even \cite{python_format,python_round,goldberg1991every}. }

\begin{threeparttable}
\begin{tabular}{ccccccccc}
\hline
\multicolumn{3}{c}{} & \multicolumn{3}{c}{0.995} & \multicolumn{3}{c}{0.9995} \\
$n$ &  $n_\G$ &  $M_n^{\mathrm{before}}$ &  $M_n^{\mathrm{after}}$  &  $s_{\mathrm{min}}$ & $ E$ & $M_n^{\mathrm{after}}$  & $s_{\mathrm{min}}$  &  $ E$   \\ \hline
1 & 10 & 21 & 17 & 2.1 $\times$ 10$^{-5}$ & -2.846763 & 20 & 9.7 $\times$ 10$^{-11}$ & -2.862423 \\
 & 14 & 29 & 23 & 9.9 $\times$ 10$^{-8}$ & -2.850469 & 27 & 5.6 $\times$ 10$^{-13}$ & -2.867865 \\
2 & 10 & 196 & 121 & 5.2 $\times$ 10$^{-9}$ & -2.883580 & 163 & 5.4 $\times$ 10$^{-17}$ & -2.896175 \\
 & 14 & 358 & 203 & 1.9 $\times$ 10$^{-11}$ & -2.876982 & 298 & 1.3 $\times$ 10$^{-17}$ & -2.900558 \\
3 & 10 & 1066 & 326 & 8.0 $\times$ 10$^{-12}$ & -2.886489 & 627 & 9.4 $\times$ 10$^{-20}$ & -2.899408 \\
 & 14 & 2444 & 560 & 4.4 $\times$ 10$^{-14}$ & -2.887513 & 1237 & 4.6 $\times$ 10$^{-23}$ & -2.902370 \\
\hline
\end{tabular} \label{tab3}
\end{threeparttable}
\end{table}

\subsection{Exponent distributions for the energy-based selection for the decontracted approach} \label{energy_sel}

In this subsection, a distribution of the exponents $\{\alpha_1, \alpha_2, \alpha_{12} \}$ via the previous decontracted approach \cite{wang2025variational} is presented. We implemented the energy selection in Refs. \cite{nakashima2020free,nakatsuji2020solving}, with using the existing Hamiltonian and overlap matrices \cite{nakashima2020free} in comparing the effect of adding new complement function, without the $O(m^2)$ updating algorithm \cite{nakashima2013efficient,nakashima2020free}. This allows us to envision the reduction of the number of $n_\G$ in the present hierarchy decontraction scheme within a certain accuracy.

As presented in Table \ref{tab4}, $37/55 \approx 67 \% $ complement functions are left after the energy-based selection without the presence of any $g$ function. With the geminal $g$ function, $21/51 \approx 42 \%$ complement function left. This is lower than the from the initial wavefunction.

\begin{figure}[H]
  \centering
  \includegraphics[width=13.5cm]{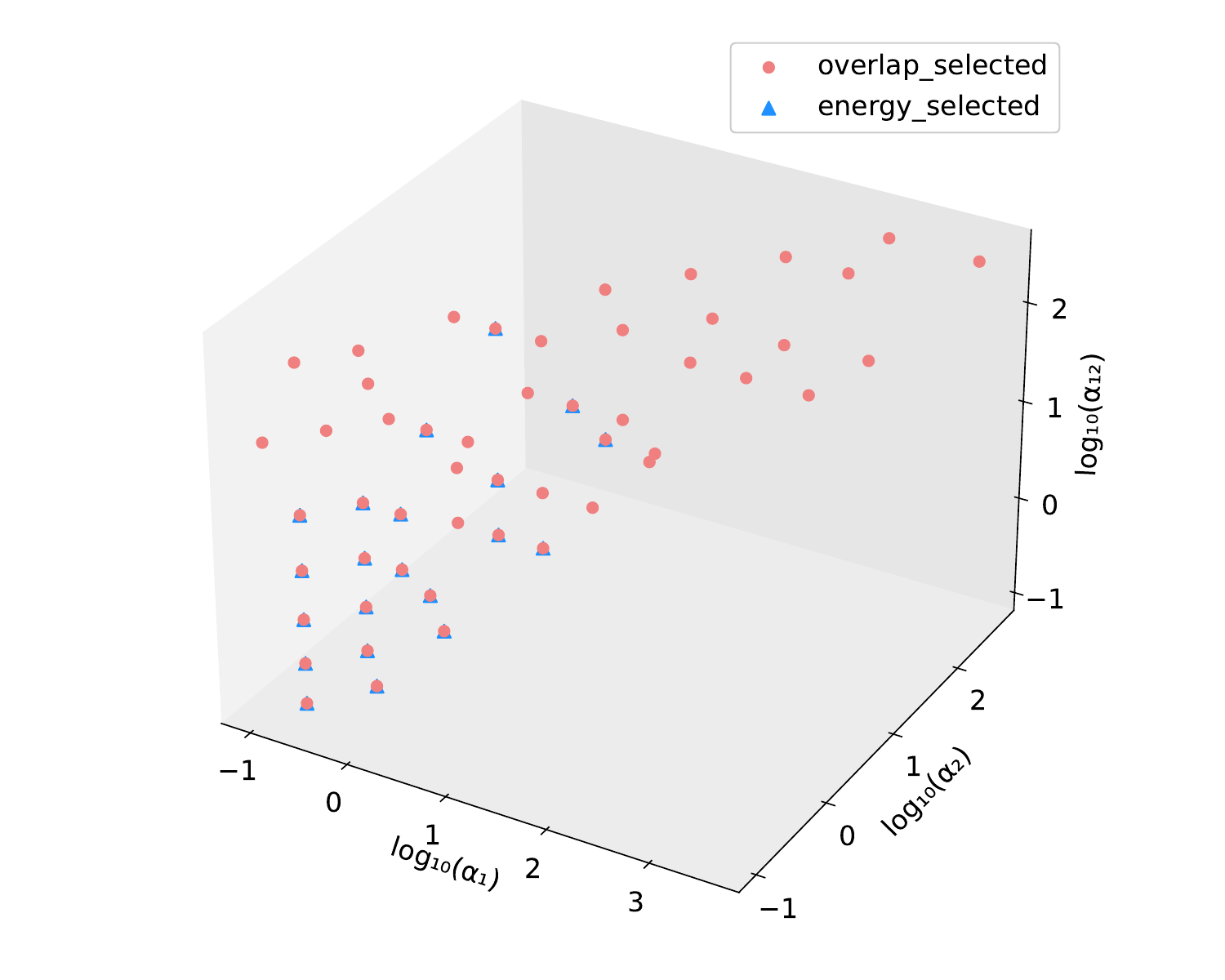}
  \caption{Distributions of the exponents $\{ \alpha_1, \alpha_2, \alpha_{12} \}$ for the helium ground state with the decontracted Gaussian-expanded FC method \cite{wang2025variational} at $n=1$ and $n_\G=10$ level, after the overlap  \cite{wang2025variational} and energy selections \cite{nakashima2020free,nakatsuji2020solving}. The thresholds of the overlap and energy selections are 0.95 and $1 \times 10^{-6}$, respectively.
  Parameters $\zeta = 1.6875 $ and $\gamma_1 = \gamma_2 = 0.3125, \gamma_{12} = 0.5$ are used with Eqs. \eqref{initial_wf} and \eqref{g_function_adopted}, respectively \cite{nakatsuji2022accurate}.
  Exponents before the symmetrizations  from the Gaussian expansion Eq. \eqref{gauss_expansion} of the complement functions associated to $n_{12} = 1$  in expression  \eqref{cf_form}  are plotted. }
  \label{fig1}
\end{figure}

Furthermore, according to Figure \ref{fig1}, fewer $\{  \alpha_1, \alpha_2 \}$ exponents after the energy-based selection compared with the overlap selection exponents from $n_1=0, n_2=0, n_{12}=1$ in expression \eqref{cf_form}. In the present hierarchical decontraction approach, this level introduced loops over the contracted Gaussians and geminal functions. This suggests a smaller number $n_\G$ compared to $n_1=0, n_2=0, n_{12}=0$ in expression \eqref{cf_form} would be necessary to achieve a certain level of accuracy. The reason may be attributed to the repulsive character of the electron-electron cusp condition \cite{kato1957eigenfunctions,pack1966cusp} and hence less energy contribution \cite{nakatsuji2020solving}. A reduction of prefactor may be anticipated for multi-electron interaction terms, e.g., 8-fold loops mentioned in Subsection \ref{slater_initial}.

Notice at $n_1 = 1, n_2 = 0, n_{12} =0 $ in expression \eqref{cf_form}, all complement functions are screened out by the overlap selection. This may due to $\gamma_1 = 0.3125$ \cite{nakatsuji2022accurate} is a relatively small number. At the level $n=2$, from $n_1 = 2, n_2 = 0, n_{12} =0 $ level with the same initial wavefunction, $g$ functions, and $n_\G=10$ in Table \ref{tab4}, 26 complement functions would be left after the overlap selection.

\begin{table}[H]
\centering
\caption{Numbers of complement functions from the decontracted Gaussian-expanded complement function method \cite{wang2025variational}. Parameters $\zeta = 1.6875 $ and $\gamma_1 = \gamma_2 = 0.3125, \gamma_{12} = 0.5$ are used with Eqs. \eqref{initial_wf} and \eqref{g_function_adopted}, respectively \cite{nakatsuji2022accurate}. $n=1$, $n_\G =10$ expansion order with decontracting the STO-10G basis set \cite{o1966gaussian} and 0.95 overlap selection threshold \cite{mitroy2013theory} are adopted. $n_1, n_2$, and $n_{12}$ correspond to the indices in expression  \eqref{cf_form}.  $M_n^{\mathrm{before}}$, $M_n^{\mathrm{after, overlap}}$, and $M_n^{\mathrm{after, energy}}$ stand for the number of complement functions before the overlap selection, after the overlap selection, and after the energy selection (the energy selection \cite{nakashima2020free,nakatsuji2020solving} follows the overlap selection), respectively. $1 \times 10^{-6}$ energy threshold was used in the energy-based selection. -2.903614 ground state energy was obtained by this energy selection.   }
\begin{tabular}{ccccc} 
\hline
$n_1, n_2, n_{12}$ & $M_n^{\mathrm{before}}$ & $M_n^{\mathrm{after, overlap}}$ & $M_n^{\mathrm{after, energy}}$ \\
\hline
$0,0,0$ & 55 & 55 & 37 \\
$1,0,0$ & 100  & 0  &  0  \\
$0,0,1$ & 550 &  51  &  21 \\
\hline
\end{tabular} \label{tab4}
\end{table}

\section{Summary and outlook}  \label{summary}
In the present work, we introduced a hierarchical decontracted Gaussian-expanded complement method in the FC framework. The decontractions are made via the Gaussian expansion, Eq. \eqref{gauss_expansion}, of the distinct, different from the initial wavefunction, Slater exponents introduced by the $g$ functions. Hence, the exponential number of variational coefficients with respect to increase the number of electrons in the previous approach \cite{wang2025variational} is postponed to higher order of the FC expansion.

We demonstrated the accuracy using the helium ground state as the example and discussed the strategies in extending into more general systems.

We noticed from the energy selection \cite{nakashima2020free,nakatsuji2020solving} combined with the previous decontracted approach \cite{wang2025variational}, a fewer number of $n_\G$ than of the initial wavefunction may be needed to obtain a certain accuracy of the energy. This suggests a reduction of computational cost for many-electron systems with the multi-loop terms.

In addition, the present method is originated from the FC method with $1-e^{-\gamma r} $ type $g$ function \cite{nakatsuji2022accurate,nakatsuji2024accurate,nakatsuji2024exact}. The Gaussian expansion, Eq. \eqref{gauss_expansion}, can be applied to any definite angular momentum and radial Slater function. For a combination of radial powers, e.g., $(1 - r + r^2) e^{-r/3}$ \cite{kurokawa2020solving}, the scaling relation, Eq. \eqref{scaling}, will not hold. Nevertheless, one can decontract this type of wavefunction. Thus, the present approach is generally applicable to many-electron systems.

\bibliographystyle{pccp}
\bibliography{topic}

\begin{thebibliography}{100}

\bibitem{shiozaki2009higher}
T.~Shiozaki, M.~Kamiya, S.~Hirata, and E.~F. Valeev, {\em J. Chem. Phys.},
  2009, {\bf 130}, 054101.

\bibitem{Valeev_PRF_2009_46811G6}
E.~Valeev, Explicitly correlated electronic structure methods for predictive
  energetics and kinetics of radical reactions ACS Petroleum Research Fund
  (PRF) 54th Annual Report on Research 2009, Report 46811-G6, 2009.

\bibitem{bubin2013born}
S.~Bubin, M.~Pavanello, W.-C. Tung, K.~L. Sharkey, and L.~Adamowicz, {\em Chem.
  Rev.}, 2013, {\bf 113}, 36--79.

\bibitem{mitroy2013theory}
J.~Mitroy, S.~Bubin, W.~Horiuchi, Y.~Suzuki, L.~Adamowicz, W.~Cencek,
  K.~Szalewicz, J.~Komasa, D.~Blume, and K.~Varga, {\em Rev. Mod. Phys.}, 2013,
  {\bf 85}, 693--749.

\bibitem{PhysRevA.95.062510}
K.~Pachucki, V.~Patk{\'o}{\v{s}}, and V.~A. Yerokhin, {\em Phys. Rev. A}, 2017,
  {\bf 95}, 062510.

\bibitem{henson2022measurement}
B.~M. Henson, J.~A. Ross, K.~F. Thomas, C.~N. Kuhn, D.~K. Shin, S.~S. Hodgman,
  Y.-H. Zhang, L.-Y. Tang, G.~W.~F. Drake, A.~T. Bondy, A.~G. Truscott, and
  K.~G.~H. Baldwin, {\em Science}, 2022, {\bf 376}, 199--203.

\bibitem{nakatsuji2004scaled}
H.~Nakatsuji, {\em Phys. Rev. Lett.}, 2004, {\bf 93}, 030403.

\bibitem{nakatsuji2005general}
H.~Nakatsuji, {\em Phys. Rev. A}, 2005, {\bf 72}, 062110.

\bibitem{nakatsuji2012discovery}
H.~Nakatsuji, {\em Acc. Chem. Res.}, 2012, {\bf 45}, 1480--1490.

\bibitem{wang2025variational}
C.~Wang, {\em arXiv preprint arXiv:2508.04635v2 [physics.chem-ph]}, 2025.

\bibitem{nakatsuji2015solving}
H.~Nakatsuji and H.~Nakashima, {\em J. Chem. Phys.}, 2015, {\bf 142}, 194101.

\bibitem{nakashima2020free}
H.~Nakashima and H.~Nakatsuji, {\em Phys. Rev. A}, 2020, {\bf 102}, 052835.

\bibitem{o1966gaussian}
K.~O-ohata, H.~Taketa, and S.~Huzinaga, {\em J. Phys. Soc. Jpn.}, 1966, {\bf
  21}, 2306--2313.

\bibitem{hehre1969self}
W.~J. Hehre, R.~F. Stewart, and J.~A. Pople, {\em J. Chem. Phys.}, 1969, {\bf
  51}, 2657--2664.

\bibitem{hehre1970self}
W.~Hehre, R.~Ditchfield, R.~Stewart, and J.~Pople, {\em J. Chem. Phys.}, 1970,
  {\bf 52}, 2769--2773.

\bibitem{pietro1980molecular}
W.~J. Pietro, B.~A. Levi, W.~J. Hehre, and R.~F. Stewart, {\em Inorg. Chem.},
  1980, {\bf 19}, 2225--2229.

\bibitem{pietro1981molecular}
W.~J. Pietro, E.~S. Blurock, R.~F. Hout~Jr, W.~J. Hehre, D.~J. DeFrees, and
  R.~F. Stewart, {\em Inorg. Chem.}, 1981, {\bf 20}, 3650--3654.

\bibitem{pietro1983molecular}
W.~J. Pietro and W.~J. Hehre, {\em J. Comput. Chem.}, 1983, {\bf 4}, 241--251.

\bibitem{lopez1987large}
R.~L{\'o}pez, G.~Ram{\'\i}rez, J.~G. de~la Vega, and J.~F. Rico, {\em J. Chim.
  Phys.}, 1987, {\bf 84}, 695--698.

\bibitem{fernandez1988accurate}
J.~Fern{\'a}ndez~Rico, G.~Ram{\'\i}rez, R.~L{\'o}pez, and J.~I.
  Fern{\'a}ndez-Alonso, {\em Collect. Czechoslov. Chem. Commun.}, 1988, {\bf
  53}, 2250--2265.

\bibitem{tew2005new}
D.~P. Tew and W.~Klopper, {\em J. Chem. Phys.}, 2005, {\bf 123}, 074101.

\bibitem{werner2007general}
H.-J. Werner, T.~B. Adler, and F.~R. Manby, {\em J. Chem. Phys.}, 2007, {\bf
  126}, 164102.

\bibitem{von2008trapped}
J.~von Stecher {\em Trapped ultracold atoms with tunable interactions} PhD
  thesis, University of Colorado at Boulder, 2008.

\bibitem{von2009correlated}
J.~von Stecher and C.~H. Greene, {\em Phys. Rev. A}, 2009, {\bf 80}, 022504.

\bibitem{rakshit2012hyperspherical}
D.~Rakshit and D.~Blume, {\em Phys. Rev. A}, 2012, {\bf 86}, 062513.

\bibitem{Kalaee2014}
A.~A.~S. Kalaee, Deuteron photodisintegration using quasicontinuum of p-waves
  in correlated gaussian basis B{Sc} thesis, Aarhus University, 2014.

\bibitem{Mosegaard2018}
P.~H. Mosegaard, Deuteron photodisintegration in a shifted correlated gaussian
  basis B{Sc} thesis, Aarhus University, 2018.

\bibitem{moriya2023novel}
H.~Moriya, W.~Horiuchi, and B.~Zhou, {\em Eur. Phys. J. A}, 2023, {\bf 59},
  197.

\bibitem{coomar2022quantum}
A.~Coomar, K.~Jones, and L.~Adamowicz, {\em Chem. Phys. Lett.}, 2022, {\bf
  790}, 139358.

\bibitem{feynman1982simulating}
R.~P. Feynman, {\em Int. J. Theor. Phys.}, 1982, {\bf 21}, 467--488.

\bibitem{aspuru2005simulated}
A.~Aspuru-Guzik, A.~D. Dutoi, P.~J. Love, and M.~Head-Gordon, {\em Science},
  2005, {\bf 309}, 1704--1707.

\bibitem{mcardle2020quantum}
S.~McArdle, S.~Endo, A.~Aspuru-Guzik, S.~C. Benjamin, and X.~Yuan, {\em Rev.
  Mod. Phys.}, 2020, {\bf 92}, 015003.

\bibitem{barca2017three}
G.~M. Barca and P.-F. Loos, {\em J. Chem. Phys.}, 2017, {\bf 147}, 024103.

\bibitem{thompson2019integral}
T.~H. Thompson and C.~Ochsenfeld, {\em J. Chem. Phys.}, 2019, {\bf 150},
  044101.

\bibitem{huron1973iterative}
B.~Huron, J.~Malrieu, and P.~Rancurel, {\em J. Chem. Phys.}, 1973, {\bf 58},
  5745--5759.

\bibitem{holmes2016heat}
A.~A. Holmes, N.~M. Tubman, and C.~J. Umrigar, {\em J. Chem. Theory Comput.},
  2016, {\bf 12}, 3674--3680.

\bibitem{tubman2016deterministic}
N.~M. Tubman, J.~Lee, T.~Y. Takeshita, M.~Head-Gordon, and K.~B. Whaley, {\em
  J. Chem. Phys.}, 2016, {\bf 145}, 044112.

\bibitem{pineda2021chembot}
S.~D. Pineda~Flores, {\em J. Chem. Theory Comput.}, 2021, {\bf 17}, 4028--4038.

\bibitem{schmerwitz2025neural}
Y.~L. Schmerwitz, L.~Thirion, G.~Levi, E.~O. Jonsson, P.~Bilous,
  H.~J{\'o}nsson, and P.~Hansmann, {\em J. Chem. Theory Comput.}, 2025, {\bf
  21}, 2301--2310.

\bibitem{casier2026machine}
B.~Casier, M.~El~Hamdi, and B.~Herzog, {\em J. Chem. Theory Comput.}, 2026,
  {\bf 22}, 1664–1673.

\bibitem{pfau2020ab}
D.~Pfau, J.~S. Spencer, A.~G. Matthews, and W.~M.~C. Foulkes, {\em Phys. Rev.
  Res.}, 2020, {\bf 2}, 033429.

\bibitem{corzo2021learning}
H.~H. Corzo, A.~Sehanobish, and O.~Kara, {\em arXiv preprint arXiv:2106.08138v2
  [quant-ph]}, 2021.

\bibitem{verstraete2023density}
F.~Verstraete, T.~Nishino, U.~Schollw{\"o}ck, M.~C. Ba{\~n}uls, G.~K. Chan, and
  M.~E. Stoudenmire, {\em Nat. Rev. Phys.}, 2023, {\bf 5}, 273--276.

\bibitem{white2023early}
S.~R. White, {\em Nat. Rev. Phys.}, 2023, {\bf 5}, 264--264.

\bibitem{baiardi2020transcorrelated}
A.~Baiardi and M.~Reiher, {\em J. Chem. Phys.}, 2020, {\bf 153}, 164115.

\bibitem{nakatsuji2022accurate}
H.~Nakatsuji, H.~Nakashima, and Y.~I. Kurokawa, {\em J. Chem. Phys.}, 2022,
  {\bf 156}, 014113.

\bibitem{nakatsuji2024accurate}
H.~Nakatsuji and H.~Nakashima, {\em J. Chem. Theory Comput.}, 2024, {\bf 20},
  3749--3765.

\bibitem{nakatsuji2009does}
H.~Nakatsuji and H.~Nakashima, {\em Int. J. Quantum Chem.}, 2009, {\bf 109},
  2248--2262.

\bibitem{klopper1986gaussian}
W.~Klopper and W.~Kutzelnigg, {\em J. Mol. Struct.: THEOCHEM}, 1986, {\bf 135},
  339--356.

\bibitem{kutzelnigg1994theory}
W.~Kutzelnigg, {\em Int. J. Quantum Chem.}, 1994, {\bf 51}, 447--463.

\bibitem{kutzelnigg1996convergence}
W.~Kutzelnigg in {\em Strategies and Applications in Quantum Chemistry: From
  Molecular Astrophysics to Molecular Engineering}, ed. Y.~Ellinger and
  M.~Defranceschi;
\newblock Springer, 1996;
\newblock pp. 79--101.

\bibitem{Kutzelnigg2011OWR}
W.~Kutzelnigg, {\em Oberwolfach Reports}, 2011, {\bf 8}, 1775--1784.

\bibitem{kutzelnigg2012rate}
W.~Kutzelnigg In {\em AIP Conf. Proc.}, Vol.  1504, pp. 15--30. American
  Institute of Physics, 2012.

\bibitem{bakken2004expansion}
V.~Bakken and T.~Helgaker, {\em Theor. Chem. Acc.}, 2004, {\bf 112}, 124--134.

\bibitem{mckemmish2012gaussian}
L.~K. McKemmish and P.~M. Gill, {\em J. Chem. Theory Comput.}, 2012, {\bf 8},
  4891--4898.

\bibitem{kutzelnigg2013expansion}
W.~Kutzelnigg, {\em Int. J. Quantum Chem.}, 2013, {\bf 113}, 203--217.

\bibitem{bachmayr2014error}
M.~Bachmayr, H.~Chen, and R.~Schneider, {\em Numer. Math.}, 2014, {\bf 128},
  137--165.

\bibitem{shaw2020completeness}
R.~A. Shaw, {\em Int. J. Quantum Chem.}, 2020, {\bf 120}, e26264.

\bibitem{wang2013rates}
C.~Wang, {\em Phys. Rev. A}, 2013, {\bf 88}, 032511.

\bibitem{morrell1975calculation}
M.~M. Morrell, R.~G. Parr, and M.~Levy, {\em J. Chem. Phys.}, 1975, {\bf 62},
  549--554.

\bibitem{hoffmann1977schrodinger}
M.~Hoffmann-Ostenhof and T.~Hoffmann-Ostenhof, {\em Phys. Rev. A}, 1977, {\bf
  16}, 1782--1785.

\bibitem{hoffmann1979lower}
T.~Hoffmann-Ostenhof, {\em J. Phys. A Math. Gen.}, 1979, {\bf 12}, 1181--1187.

\bibitem{hoffmann1980lower}
T.~Hoffmann-Ostenhof, {\em Phys. Lett. A}, 1980, {\bf 77}, 140--142.

\bibitem{katriel1980asymptotic}
J.~Katriel and E.~R. Davidson, {\em Proc. Natl. Acad. Sci. U.S.A.}, 1980, {\bf
  77}, 4403--4406.

\bibitem{ahlrichs1981bounds}
R.~Ahlrichs, M.~Hoffmann-Ostenhof, T.~Hoffmann-Ostenhof, and J.~D. Morgan~III,
  {\em Phys. Rev. A}, 1981, {\bf 23}, 2106--2117.

\bibitem{simon1982schrodinger}
B.~Simon, {\em Bull. Am. Math. Soc.}, 1982, {\bf 7}, 447--526.

\bibitem{froese1983exponential}
R.~Froese and I.~Herbst, {\em Commun. Math. Phys.}, 1983, {\bf 92}, 71--80.

\bibitem{ahlrichs1989basic}
R.~Ahlrichs in {\em Numerical Determination of the Electronic Structure of
  Atoms, Diatomic and Polyatomic Molecules}, ed. M.~Defranceschi and
  J.~Delhalle;
\newblock Springer, 1989;
\newblock pp. 1--15.

\bibitem{fournais2008local}
S.~Fournais, M.~Hoffmann-Ostenhof, T.~Hoffmann-Ostenhof, and T.~{\O}.
  S{\o}rensen, {\em AIP Conf. Proc.}, 2008, {\bf 998}, 70--84.

\bibitem{agmon2014lectures}
S.~Agmon, {\em Lectures on Exponential Decay of Solutions of Second-Order
  Elliptic Equations: Bounds on Eigenfunctions of N-Body Schrodinger
  Operations.(MN-29)}, Princeton University Press, 2014.

\bibitem{klahn1977convergence}
B.~Klahn and W.~A. Bingel, {\em Theor. Chim. Acta}, 1977, {\bf 44}, 27--43.

\bibitem{lowdin1955quantum_a}
P.-O. L{\"o}wdin, {\em Phys. Rev.}, 1955, {\bf 97}, 1474--1489.

\bibitem{lowdin1955quantum_b}
P.-O. L{\"o}wdin, {\em Phys. Rev.}, 1955, {\bf 97}, 1490--1508.

\bibitem{nakatsuji2020solving}
H.~Nakatsuji, H.~Nakashima, and Y.~I. Kurokawa, {\em Phys. Rev. A}, 2020, {\bf
  101}, 062508.

\bibitem{Limpanuparb2011}
T.~Limpanuparb {\em Applications of Resolutions of the Coulomb Operator in
  Quantum Chemistry} PhD thesis, Australian National University, 2011.

\bibitem{parrish2013exact}
R.~M. Parrish, E.~G. Hohenstein, N.~F. Schunck, C.~D. Sherrill, and T.~J.
  Mart{\'\i}nez, {\em Phys. Rev. Lett.}, 2013, {\bf 111}, 132505.

\bibitem{khoromskaia2015tensor}
V.~Khoromskaia and B.~N. Khoromskij, {\em Phys. Chem. Chem. Phys.}, 2015, {\bf
  17}, 31491--31509.

\bibitem{khoromskaia2018tensor}
V.~Khoromskaia and B.~N. Khoromskij, {\em Tensor numerical methods in quantum
  chemistry}, Walter de Gruyter GmbH \& Co KG, 2018.

\bibitem{Glaser2024}
N.~Glaser and M.~Reiher, {\em Chimia}, 2024, {\bf 78}, 215--221.

\bibitem{kutzelnigg1991wave}
W.~Kutzelnigg and W.~Klopper, {\em J. Chem. Phys.}, 1991, {\bf 94}, 1985--2001.

\bibitem{helgaker2000molecular}
T.~Helgaker, P.~Jorgensen, and J.~Olsen, {\em Molecular electronic-structure
  theory}, John Wiley \& Sons, 2000.

\bibitem{python_math}
\url{https://docs.python.org/3/library/math.html#math.isclose}, Accessed:
  2026-03-15.

\bibitem{silvestre2007few}
B.~Silvestre-Brac and V.~Mathieu, {\em Phys. Rev. E}, 2007, {\bf 76}, 046702.

\bibitem{garniron2019quantum}
Y.~Garniron, T.~Applencourt, K.~Gasperich, A.~Benali, A.~Fert{\'e}, J.~Paquier,
  B.~Pradines, R.~Assaraf, P.~Reinhardt, J.~Toulouse, et~al., {\em J. Chem.
  Theory Comput.}, 2019, {\bf 15}, 3591--3609.

\bibitem{van1991interactively}
G.~Van~Rossum and J.~De~Boer, {\em CWI quarterly}, 1991, {\bf 4}, 283--303.

\bibitem{mpmath}
mpmath: a {P}ython library for arbitrary-precision floating-point arithmetic
  (version 1.3.0). {The mpmath development team} 2023.

\bibitem{time_python}
\url{https://docs.python.org/3.12/library/time.html}, Accessed: 2026-03-15.

\bibitem{Hunter:2007}
J.~D. Hunter, {\em Comput. Sci. Eng.}, 2007, {\bf 9}, 90--95.

\bibitem{10.7717/peerj-cs.103}
A.~Meurer, C.~P. Smith, M.~Paprocki, O.~\v{C}ert\'{i}k, S.~B. Kirpichev,
  M.~Rocklin, A.~Kumar, S.~Ivanov, J.~K. Moore, S.~Singh, T.~Rathnayake,
  S.~Vig, B.~E. Granger, R.~P. Muller, F.~Bonazzi, H.~Gupta, S.~Vats,
  F.~Johansson, F.~Pedregosa, M.~J. Curry, A.~R. Terrel, v.~Rou\v{c}ka,
  A.~Saboo, I.~Fernando, S.~Kulal, R.~Cimrman, and A.~Scopatz, {\em PeerJ
  Comput. Sci.}, 2017, {\bf 3}, e103.

\bibitem{cachetools}
\url{https://github.com/tkem/cachetools}, Accessed: 2026-05-21.

\bibitem{Julia-2017}
J.~Bezanson, A.~Edelman, S.~Karpinski, and V.~B. Shah, {\em SIAM Rev.}, 2017,
  {\bf 59}, 65--98.

\bibitem{json}
\url{https://github.com/JuliaIO/JSON.jl}, Accessed: 2026-05-21.

\bibitem{python_set}
\url{docs.python.org/3/library/stdtypes.html}, Accessed: 2026-03-15.

\bibitem{cursor}
\url{https://www.cursor.com/}, Accessed: 2026-03-15.

\bibitem{schwartz2006further}
C.~Schwartz, {\em arXiv preprint math-ph/0605018}, 2006.

\bibitem{nakashima2007solving}
H.~Nakashima and H.~Nakatsuji, {\em J. Chem. Phys.}, 2007, {\bf 127}, 224104.

\bibitem{kurokawa2008solving}
Y.~I. Kurokawa, H.~Nakashima, and H.~Nakatsuji, {\em Phys. Chem. Chem. Phys.},
  2008, {\bf 10}, 4486--4494.

\bibitem{nakashima2008accurately}
H.~Nakashima and H.~Nakatsuji, {\em Phys. Rev. Lett.}, 2008, {\bf 101}, 240406.

\bibitem{hesselmann2005efficient}
A.~He{\ss}elmann and F.~Manby, {\em J. Chem. Phys.}, 2005, {\bf 123}.

\bibitem{szabo1996modern}
A.~Szabo and N.~S. Ostlund, {\em Modern quantum chemistry: introduction to
  advanced electronic structure theory}, Dover Publications, 1996.

\bibitem{python_format}
\url{https://docs.python.org/3/library/string.html#format-specification-mini-language},
  Accessed: 2026-03-15.

\bibitem{python_round}
\url{https://docs.python.org/3/library/functions.html#round}, Accessed:
  2026-03-15.

\bibitem{goldberg1991every}
D.~Goldberg, {\em ACM Comput. Surv.}, 1991, {\bf 23}, 5--48.

\bibitem{kato1957eigenfunctions}
T.~Kato, {\em Commun. Pure Appl. Math.}, 1957, {\bf 10}, 151--177.

\bibitem{pack1966cusp}
R.~T. Pack and W.~B. Brown, {\em J. Chem. Phys.}, 1966, {\bf 45}, 556--559.

\bibitem{nakashima2013efficient}
H.~Nakashima and H.~Nakatsuji, {\em J. Chem. Phys.}, 2013, {\bf 139}, 044112.

\bibitem{nakatsuji2024exact}
H.~Nakatsuji and H.~Nakashima, {\em J. Chem. Theory Comput.}, 2024, {\bf 20},
  8001--8009.

\bibitem{kurokawa2020solving}
Y.~I. Kurokawa, H.~Nakashima, and H.~Nakatsuji, {\em Phys. Chem. Chem. Phys.},
  2020, {\bf 22}, 13489--13497.

\end{thebibliography}

\end{document}